\documentclass[twocolumn]{aastex701}

\newcommand{\oiiiuv}{O~III] $\lambda 1666$}
\newcommand{\ciii}{C~III] $\lambda 1908$}
\newcommand{\civ}{C~IV $\lambda 1549$}
\newcommand{\heii}{He~II $\lambda 1640$}

\newcommand{\niv}{N~IV] $\lambda 1488$}
\newcommand{\mgii}{Mg~II] $\lambda 2880$}
\newcommand{\nev}{[Ne~V] $\lambda 3426$}

\newcommand{\oiiiboth}{[O~III] $\lambda\lambda 4959, 5007$}

\begin{document}

\title{Spectroscopic confirmation of a large and luminous galaxy with weak emission lines at $\mathbf{z = 13.53}$}

\author[0000-0002-7622-0208]{Callum T. Donnan}
\affiliation{NSF's National Optical-Infrared Astronomy Research Laboratory, 950 N. Cherry Ave., Tucson, AZ 85719, USA}
\email[show]{callum.donnan@noirlab.edu}

\author[0000-0003-4368-3326]{Derek J. McLeod}
\affiliation{Institute for Astronomy, University of Edinburgh, Royal Observatory, Edinburgh EH9 3HJ, UK}
\email{derek.mcleod@ed.ac.uk}

\author[0009-0005-9742-2318]{Ross J. McLure}
\affiliation{Institute for Astronomy, University of Edinburgh, Royal Observatory, Edinburgh EH9 3HJ, UK}
\email{rmclure@ed.ac.uk}

\author[0000-0002-1404-5950]{James S. Dunlop}
\affiliation{Institute for Astronomy, University of Edinburgh, Royal Observatory, Edinburgh EH9 3HJ, UK}
\email{james.dunlop@ed.ac.uk}

\author[0000-0002-3736-476X]{Fergus Cullen}
\affiliation{Institute for Astronomy, University of Edinburgh, Royal Observatory, Edinburgh EH9 3HJ, UK}
\email{fergus.cullen@ed.ac.uk}

\author[0000-0001-5414-5131]{Mark Dickinson}
\affiliation{NSF's National Optical-Infrared Astronomy Research Laboratory, 950 N. Cherry Ave., Tucson, AZ 85719, USA}
\email{mark.dickinson@noirlab.edu}

\author[0000-0002-7959-8783]{Pablo Arrabal Haro}
\affiliation{Center for Space Sciences and Technology, UMBC, 5523 Research Park Dr, Baltimore, MD 21228 USA }
\affiliation{Astrophysics Science Division, NASA Goddard Space Flight Center, 8800 Greenbelt Rd, Greenbelt, MD 20771, USA}
\email{pablo.arrabalharo@nasa.gov }

\author[0000-0003-1282-7454]{Anthony J. Taylor}
\email{anthony.taylor@austin.utexas.edu}
\affiliation{Department of Astronomy, The University of Texas at Austin, Austin, TX, USA}
\affiliation{Cosmic Frontier Center, The University of Texas at Austin, Austin, TX, USA}

\author[0009-0008-3775-5112]{Cecilia Bondestam}
\affiliation{Institute for Astronomy, University of Edinburgh, Royal Observatory, Edinburgh EH9 3HJ, UK}
\email{C.Bondestam@ed.ac.uk}

\author[0000-0003-1386-3676]{Feng-Yuan Liu}
\affiliation{Institute for Astronomy, University of Edinburgh, Royal Observatory, Edinburgh EH9 3HJ, UK}
\email{Fengyuan.Liu@ed.ac.uk }

\author[0000-0002-2644-3518]{Karla Z. Arellano-C\'{o}rdova}
\affiliation{Institute for Astronomy, University of Edinburgh, Royal Observatory, Edinburgh EH9 3HJ, UK}
\email{K.Arellano@ed.ac.uk}

\author[0000-0003-1641-6185]{Laia Barrufet}
\affiliation{Institute for Astronomy, University of Edinburgh, Royal Observatory, Edinburgh EH9 3HJ, UK}
\email{lbarrufe@roe.ac.uk }

\author[0000-0003-0629-8074]{Ryan Begley}
\affiliation{{Armagh Observatory and Planetarium, College Hill, Armagh, BT61 9DG, N. Ireland, UK}}
\email{ryan.begley@armagh.ac.uk}

\author[0000-0002-1482-5818]{Adam C. Carnall}
\affiliation{Institute for Astronomy, University of Edinburgh, Royal Observatory, Edinburgh EH9 3HJ, UK}
\email{adamc@roe.ac.uk }

\author[0009-0005-9306-9633]{Hanna Golawska}
\affiliation{Institute for Astronomy, University of Edinburgh, Royal Observatory, Edinburgh EH9 3HJ, UK}
\email{H.Golawska@sms.ed.ac.uk }

\author[0000-0003-0486-5178]{Ho-Hin Leung}
\affiliation{Institute for Astronomy, University of Edinburgh, Royal Observatory, Edinburgh EH9 3HJ, UK}
\email{hleung2@roe.ac.uk}

\author[0000-0002-6867-1244]{Dirk Scholte}
\affiliation{Institute for Astronomy, University of Edinburgh, Royal Observatory, Edinburgh EH9 3HJ, UK}
\email{dscholte@roe.ac.uk}

\author[0000-0002-0827-9769]{Thomas M. Stanton}
\affiliation{Institute for Astronomy, University of Edinburgh, Royal Observatory, Edinburgh EH9 3HJ, UK}
\email{T.Stanton@ed.ac.uk}

\begin{abstract}
We present \textit{JWST}/NIRSpec PRISM observations of a robust galaxy candidate at $z\simeq14$, selected from pure-parallel NIRCam imaging; PAN-z14-1. The NIRSpec spectrum allows confirmation of this source at $z_{\rm spec}=13.53^{+0.05}_{-0.06}$ through modeling of the Lyman-$\alpha$ break. PAN-z14-1 is the fourth
most distant galaxy known to date and is extremely luminous ($M_{\rm UV}=-20.6\pm0.2$), with a blue UV-continuum slope ($\beta=-2.26\pm0.08$) and a large physical size ($r_{\rm c}=233\pm10\, \rm pc$). 
We fail to detect any rest-frame UV emission lines at $\geq 2\sigma$ significance, with upper limits sufficiently constraining to exclude the possibility of strong line emission. In terms of its physical properties, PAN-z14-1 is remarkably similar to the previously confirmed $z_{\rm spec}=14.18$ galaxy GS-z14-0. 
The lack of strong emission lines and large physical size is consistent with an emerging picture of two potentially distinct galaxy populations at $z>10$, distinguished by star-formation rate surface density.
In this scenario, PAN-z14-1 is a second example of a ``normal'', extended, luminous, star-forming galaxy at $z \simeq 14$, and differs markedly from the other class of extremely compact galaxies with strong emission lines recently uncovered at extreme redshifts with {\it JWST}.
These results highlight the importance of further spectroscopic confirmation of $z>10$ galaxy candidates in order to fully understand the diversity of properties displayed by the first galaxies. 
\end{abstract}



\section{Introduction}
One of the key goals of \textit{JWST} is to detect and characterize the first galaxies which formed in the early history of the Universe. Prior to the launch of \textit{JWST}, galaxies had been detected out to redshifts $z\simeq10$ using both \textit{HST} and \textit{Spitzer}, as well as wide-area ground-based surveys \citep[e.g.,][]{ellis2013,mclure2013,oesch2014,oesch2018,bowler2014,bowler2015,bowler2020,finkelstein2015,mcleod2015,mcleod2016,bouwens2021,bouwens2022}. Since its launch, the sensitive imaging at $\lambda=1-5\, \mu$m provided by \textit{JWST} has produced large samples of robust galaxy candidates at $z>10$ for the first time \citep[e.g.,][]{castellano2022,naidu2022,donnan2023b, donnan2023a,adams2022, finkelstein2022c,harikane2023a,franco2024}. 
These samples have been used to constrain the galaxy UV luminosity function (LF) at $z\simeq9-14$ \citep[e.g.,][]{finkelstein2024, perezgonzalez2023,2025arXiv250315594P, mcleod2024, adams2023, donnan2024,whitler2025} with a general consensus emerging that evolution of the UV LF is slower than predicted by many pre-\textit{JWST} theoretical models \citep{mason2015,tacchella2018,yung2019}, but consistent with some pre-\textit{JWST} observations at $z=9-10$ \citep{mcleod2016}. 

However, at $z>14$ the evolution of the galaxy UV LF remains significantly more uncertain. If confirmed, recent reports of {\it potential} candidates at $z=18-25$ \citep[e.g.][]{kokorev2025b,castellano2025,perezgonzalez2025} would indicate surprisingly high galaxy number densities at $z>14$. This is in direct contrast to the more rapid decline in the number density of galaxies at $z\geq14$ determined by \citet{donnan2024,donnan2025}, a picture that is supported by the recent, wide-area, \textit{JWST} studies of
\citet{Weibel2025} and D.\,J.\,McLeod et al. (in prep).

The most direct method to resolve this apparent tension is the spectroscopic verification of galaxy candidates at $z \geq 14$. \textit{JWST}/NIRSpec has now been successfully exploited to confirm the redshifts of many galaxy candidates at $z\geq10$ \citep[e.g.,][]{curtislake2022, bunker2023, arrabalharo2023, carniani2024, castellano2024,napolitano2024, napolitano2025, kokorev2025}, with the highest-redshift, spectroscopically-confirmed galaxy currently at $z_{\rm spec}=14.44$ \citep{naidu2025}.

These spectroscopic data have also already revealed that $z\geq 10$ galaxies display a variety of physical properties. For example, while the majority appear to be dust-poor with blue UV-continuum slopes \citep{robertsborsani2025}, a few are red enough to suggest significant dust attenuation \citep{donnan2025b,tang2025}. Moreover, there is also diversity in the strength of their rest-frame UV emission lines, with some galaxies showing strong carbon and nitrogen lines, indicating elevated nitrogen and sub-solar carbon abundances \citep[e.g.,][]{bunker2023,castellano2024}, and others showing evidence of active galactic nuclei (AGN) activity \citep[e.g.][]{napolitano2024,taylor2025}. In contrast, 
many other $z\geq10$ galaxies display weak (or at least non-detected) UV emission lines \citep[e.g.,][]{curtislake2022,carniani2024}. 
Understanding the cause of this diversity in rest-frame UV properties and chemical abundances (in particular their nitrogen and carbon abundances) remains a challenge. Interestingly, many of the strong line-emitters appear to have compact morphologies \citep{harikane2025,robertsborsani2025}, which could suggest that their UV line strength is linked to high star-formation surface density \citep{schaerer2024}. 

To help resolve the significant tension in the abundance of galaxies at $z\geq14$, as well as to further characterize their properties, we present \textit{JWST}/NIRSpec observations of a robust $z\simeq14$ candidate selected from NIRCam imaging data; PAN-z14-1. This galaxy was first reported as a $z\simeq14.5$ candidate by \citet{Golawska2024} as part of a search for $z>12$ galaxies in the PANORAMIC survey (GO 2514; PIs Williams, Oesch; see \citealt{williams2025}), before being re-selected as a $z\simeq14.4$ candidate in a wide area search ($\gtrsim0.5$ sq deg) for $z>12$ galaxies, based on a compilation of publicly available \textit{JWST} ERS, ERO and Cycles 1-4 NIRCam surveys (McLeod et al. in prep). It has also been reported as a $z\simeq 14.5$ candidate in \citet{Weibel2025}. Here we present and analyse new NIRSpec PRISM observations obtained in {\it JWST} Cycle-4 (GO 6954, PIs: Donnan, McLeod) which were specifically designed to verify the redshift of this source and measure its physical properties.

This paper is structured as follows. In Section~\ref{sec:data} we describe the photometric and spectroscopic data used in this work. In Section~\ref{sec:analysis} we describe the redshift fitting, spectral energy distribution (SED) modeling, emission-line fitting and morphological measurements. In Section~\ref{sec:discussion} we discuss the physical nature of PAN-z14-1. Finally, in Section~\ref{sec:conclusion} we summarize our conclusions. Throughout we use magnitudes in the AB system \citep{oke1974,oke1983}, adopt a \citet{Kroupa2001} initial mass function (IMF), and assume a standard cosmological model with $H_0=70$\,km\,s$^{-1}$\,Mpc$^{-1}$, $\Omega_m=0.3$ and $\Omega_{\Lambda}=0.7$.  

\section{Data}
\label{sec:data}
In this section, we provide a summary of the \textit{JWST} NIRSpec and NIRCam data sets used in this study.
\subsection{NIRSpec data}
\label{sec:data_nirspec}
Our \textit{JWST}/NIRSpec data were obtained through the Cycle-4 GO program 6954 (PIs Donnan, McLeod) in fixed-slit mode using the PRISM with an exposure time of 14574.3 seconds (4.1 hours). The data were reduced using version 1.20.1 of the NIRSpec data reduction pipeline, with CRDS version \texttt{1464.pmap}. The data reduction process follows the method described in \citet{arrabalharo2023b}, with some modifications. The \texttt{clean\_flicker\_noise} option is used to remove the $1/f$ noise at the \texttt{calwebb\_detector1} stage. In \texttt{calwebb\_spec2} we also use a modified flat-field and, after running the \texttt{calwebb\_spec3} stage, we obtain a reduced 2D spectrum before generating the 1D spectrum using optimal extraction \citep{horne1986}. The background subtraction is performed using a nodded sky-subtraction at the second stage of the reduction pipeline. The 1D spectrum is corrected for slit-losses by matching the flux density to the measured photometry. This is done by using the 1D spectrum to generate synthetic fluxes in all the relevant NIRCam filters and fitting a second-order polynomial to the ratio of the photometric-to-synthetic fluxes. This is described in further detail in Appendix~\ref{sec:apdx_slit_loss}.

\subsection{NIRCam data}
The NIRCam data used in the spectro-photometric fitting is from the Cycle-1 GO program PANORAMIC, with the observations taken in November 2023. This data set was reduced with the PRIMER Enhanced NIRCam Image-processing Library (\textsc{pencil}; D.\,Magee et al. in prep). This is a customized version of the \textit{JWST} pipeline v1.17.1, featuring steps for the treatment of wisps, snowballs, 1/f striping noise and background subtraction. The CRDS context is \texttt{1414.pmap}. The NIRCam data includes imaging in F115W, F150W, F200W, F277W, F356W, F410M and F444W filters, with 5$\sigma$ limiting magnitudes of 29.1, 29.0, 29.1, 29.1, 29.0, 28.6 and 28.7 mag, respectively (0.35$^{\prime\prime}-$diameter apertures; point-source total). As the object is extended, we measure photometry in 0.35$^{\prime\prime}-$diameter apertures using \textsc{Source-Extractor} \citep{bertin1996} in dual-image mode, with the F277W data utilised as the detection image. Prior to measuring the photometry, the imaging data in all filters was PSF-homogenized to match the spatial resolution of the F444W data. We measure photometric uncertainties by first generating a grid of 0.35$^{\prime\prime}-$diameter apertures spanning the field-of-view, and then measuring the $\sigma$ from the nearest 200 blank-sky apertures to the target after masking objects with a segmentation map, following \citet{Begley2025}. We correct fluxes to total by scaling to FLUX\_AUTO using Kron apertures \citep{kron1980} and adding a further 10\% correction, following \citet{mcleod2024}.

\section{Data Analysis}
\label{sec:analysis}

\begin{deluxetable*}{lc}
\label{tab:sample}
\caption{Basic observed properties of PAN-z14-1.}
\tablehead{Property & Value}

\startdata
    ID & PAN-z14-1\\
    RA /deg & $334.25035598$\\
    DEC /deg & $0.3792145611$\\
    $z_{\rm spec}$ & $13.53^{+0.05}_{-0.06}$\\
    $M_{\mathrm{UV}}$ & $-20.6\pm0.2$\phantom{0}\\
    $\beta$ & $-2.26\pm0.08$\\
    $z_{\rm phot}$ & $14.43^{+0.39}_{-0.29}$\\
\enddata
\end{deluxetable*}

\subsection{Spectroscopic redshift}
\label{sec:spec_z}
\begin{figure*}[ht!]
\centering
\includegraphics[width=\textwidth]
{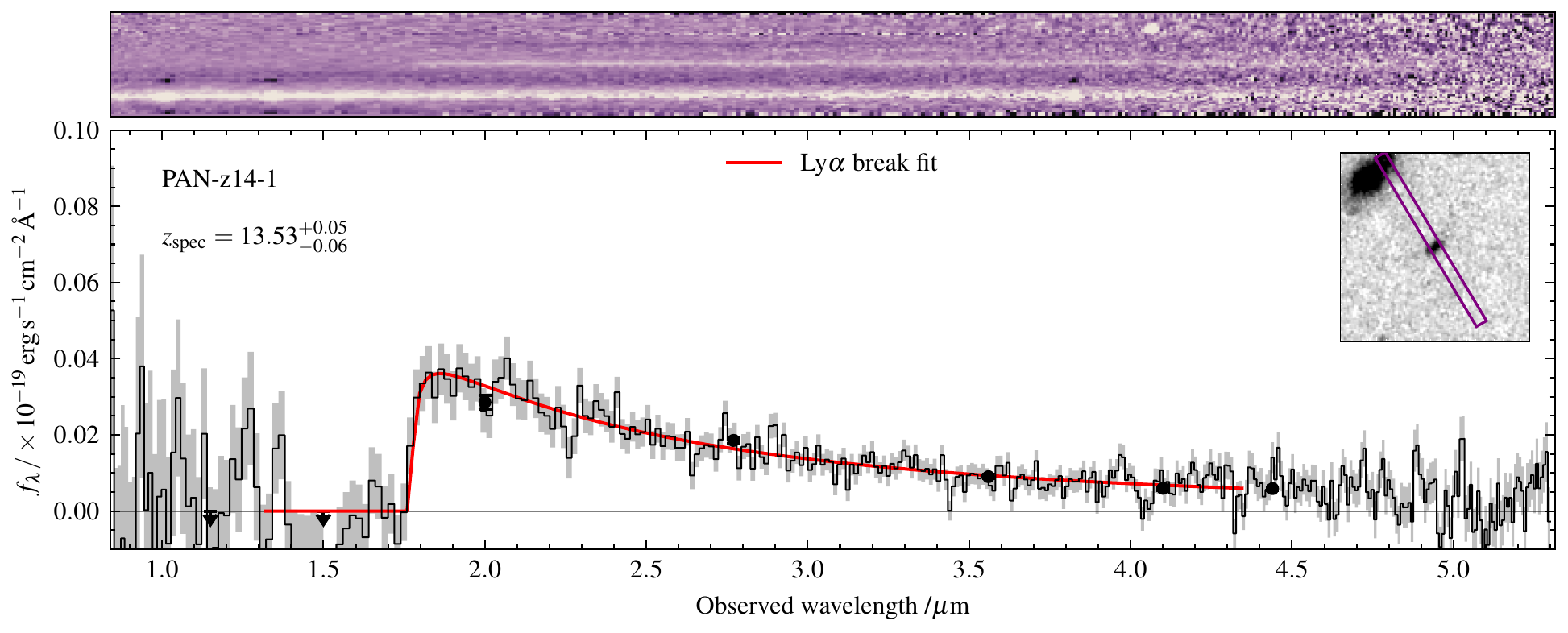}
\caption{The 2D (top) and 1D (bottom) spectra of PAN-z14-1. The black data points show the measured NIRCam photometry and the red line shows the best-fitting model of the Lyman-$\alpha$ break and UV continuum. A $3.15\times3.15$ arcsec cutout image of the source in the F277W filter is shown in the inset with the NIRSpec slit overlayed.}
\label{fig:spec}
\end{figure*}

In Fig.~\ref{fig:spec} we show the NIRSpec PRISM spectrum (2D and extracted 1D) of PAN-z14-1 which displays a definitive Lyman-$\alpha$ break at $\simeq 1.75\, \mu \rm m$, a blue UV-continuum, but also a notable lack of strong emission lines. Therefore, to obtain a spectroscopic redshift we fit a model to the Lyman-$\alpha$ break following \citet{cullen24}. This model includes three free parameters: the source redshift, the UV-continuum slope ($\beta$, where $f_{\lambda}\propto \lambda^{\beta}$) and the fraction of neutral hydrogen in the intergalactic medium ($X_{\rm HI}$), which damps the break according to the prescription from \citet{miralda98}. We assume the following flat priors for each parameter: $-10\leq \beta \leq10$, $5\leq z \leq 20$ and $0\leq \mathrm{X_{HI} \leq1}$. The fitting is further described and resultant posterior distributions are shown in Appendix~\ref{sec:apdx_z_fit}.

This yields a spectroscopic redshift $z_{\rm spec}=13.53^{+0.05}_{-0.06}$ with the best-fitting model shown by the red line in Fig.~\ref{fig:spec}. 
The successful confirmation of the high-redshift nature of PAN-z14-1, which was first selected from a sample of $z\geq12$ galaxy candidates in D. J. McLeod et al. (in prep.), demonstrates the strength of requiring an $8\sigma$ detection without the need to use a redshift prior (first used in \citealp{mcleod2024}). Note that this is different to the selection implemented in \citet{donnan2024} which use a UV luminosity function prior to accurately remove low-redshift contaminants when only requiring a $5\sigma$ detection.

Although PAN-z14-1 is confirmed to be a high-redshift galaxy, the spectroscopic redshift is somewhat lower than our original photometric redshift of $z_{\rm phot} = 14.43^{+0.39}_{-0.29}$ (D. J. McLeod et al., in prep). We discuss potential explanations for this in Section~\ref{sec:redshift_comparison}.

\subsection{UV properties}
\begin{figure*}[ht!]
\centering
\includegraphics[width=\textwidth]
{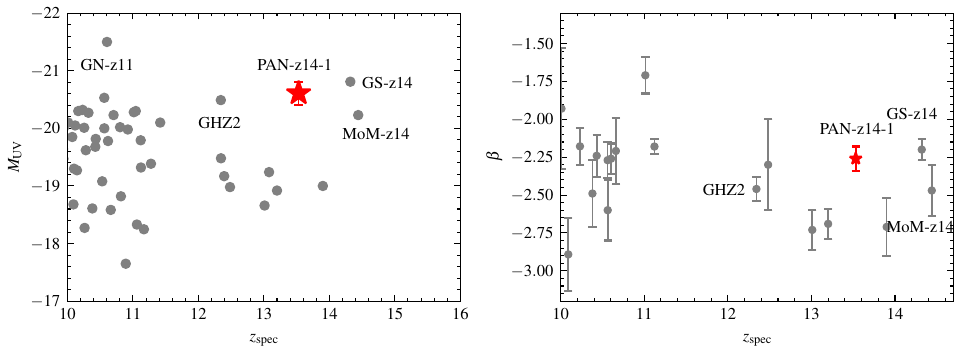}
\caption{\textit{Left}: The $M_{\rm UV}-z_{\rm spec}$ plane for spectroscopically-confirmed galaxies at $z>10$, with the location of PAN-z14-1 highlighted by the red star. \textit{Right}: The corresponding $\beta-z_{\rm spec}$ plane for spectroscopically-confirmed $z>10$ galaxies.}
\label{fig:Muv_z}
\end{figure*}

From our fit to the 1D spectrum, we also measure a UV-continuum slope of $\beta=-2.26\pm0.08$ and a neutral fraction of $X_{\rm HI}=0.75^{+0.18}_{-0.26}$, consistent with a highly-neutral IGM at $z=13.5$. We measure an absolute UV magnitude of $M_{\rm UV}=-20.6\pm0.2$ by directly integrating the photometry-corrected spectrum through a tophat filter of width $100\, \rm \AA$ centered on a rest-frame wavelength of $1500\, \rm \AA$. A comparison to other spectroscopically confirmed galaxies at $z>10$ is shown in Fig.~\ref{fig:Muv_z}, for both $M_{\rm UV}$ and $\beta$ as a function of redshift. PAN-z14-1 is currently the fourth highest-redshift spectroscopically-confirmed galaxy and exhibits a similar UV luminosity to GS-z14-0 \citep{carniani2024}, a UV luminosity only significantly exceeded at $z > 10$ by GN-z11 \citep{bunker2023b}. The UV slope is somewhat redder than most of the spectroscopically-confirmed galaxies at $z>10$, but is again very similar to that of GS-z14-0. 

\subsection{SED fitting}
\label{sec:SED_fitting}
To measure the physical properties of PAN-z14-1 we performed simultaneous SED modeling of the photometric and spectroscopic data using \textsc{bagpipes} \citep{carnall2018,carnall2019}, which accurately accounts for the variable spectral resolution of the NIRSpec PRISM data. The \textsc{bagpipes} modeling used \citet{bruzual2003} stellar population models with the MILES stellar library \citep{Falcon-Barroso2011} and a \citet{Kroupa2001} IMF. We assumed a \citet{calzetti2000} dust-attenuation law with a uniform prior on the V-band attenuation over the range: $A_{\rm V}=0-3$. We also assumed a log-uniform prior on the stellar mass ($\log(M_{\star}/\rm M_{\odot}) = 5.0-13.0$), a log-uniform prior on metallicity ($Z/ \mathrm{Z_{\odot}} = 0 - 2.5$) and a log-uniform prior on the ionization parameter ($-4<\log U<0$). 

\begin{figure*}[ht!]
\centering
\includegraphics[width=\textwidth]
{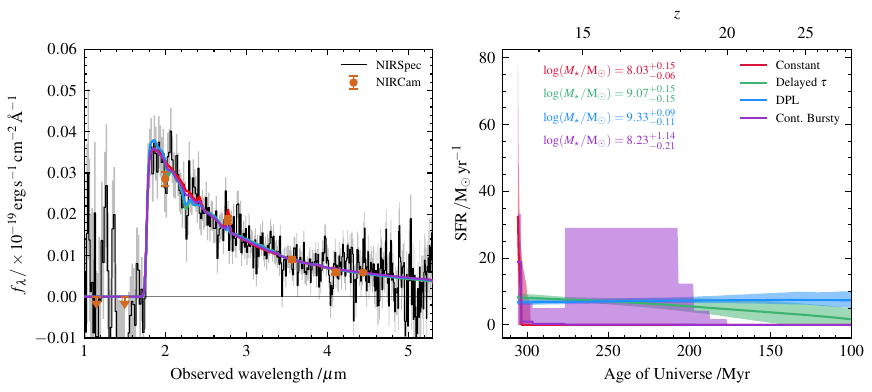}
\caption{\textit{Left}: The best-fitting spectral energy distribution (SED) models to the NIRSpec and NIRCam data for PAN-z14-1 using \textsc{bagpipes} with a constant (red), delayed-$\tau$ (green), double power-law (blue) and continuity-bursty non-parametric (purple) star-formation history (SFH) model. \textit{Right}: The corresponding SFH for each best-fitting \textsc{bagpipes} SED model. There is significant scatter ($\simeq1.3\, \rm dex$) in the inferred stellar mass depending on the choice of model for the SFH.}
\label{fig:SED_fit}
\end{figure*}

We tested a number of different star-formation history (SFH) models when fitting the NIRSpec and NIRCam data.
Firstly, we adopted a constant SFH model with a log-uniform prior on the stellar age between $1\, \rm Myr$ and $10\, \rm Gyr$. Secondly, we adopted a delayed-$\tau$ model with a uniform prior on the mass-weighted stellar age of $t_{\star} / \mathrm{Gyr}\in(0.01,5)$ and a uniform prior on $\log(\tau)\in(0.1,10)$.  Thirdly, we adopted a double power-law (DPL) model that parametrizes the SFH with rising and falling power laws. Here we imposed a log-uniform prior on each slope ($d\log(\mathrm{SFR})/d\log t$) between 0.01 and 1000. Finally, we adopted a non-parametric model of the SFH. We assumed a ``continuity-bursty'' model, as defined in \citet{tacchella2022}, that allows the SFH to change over 8 bins in cosmic time. The first 4 bins are spaced logarithmically between $z=20$ and $100\, \rm Myr$, with the final 4 bins defined between $30-100 \, \rm Myr$, $10-30 \, \rm Myr$, $3-10 \, \rm Myr$ and  $0-3 \, \rm Myr$.
The results of the SED modeling are shown in Fig.~\ref{fig:SED_fit}.

There are significant differences in the inferred stellar mass of PAN-z14-1 returned by the different SFH models. The delayed-$\tau$ and DPL models prefer an older stellar age ($t_{\star}>100\, \rm Myr$ with a SFH extending to $z>20$). This results in a high stellar mass of $\log(M_{\star}/\mathrm{M_{\odot}})=9.33^{+0.09}_{-0.11}$ for the DPL SFH and $\log(M_{\star}/\mathrm{M_{\odot}})=9.07^{+0.15}_{-0.15}$ for the delayed-$\tau$ SFH. 

In contrast, the constant SFH model prefers a significantly lower stellar mass of $\log(M_{\star}/\mathrm{M_{\odot}})=8.03^{+0.15}_{-0.06}$ and a younger mass-weighted stellar age of $t_{\star} = 2^{+3}_{-1}\, \rm Myr$. The continuity-bursty SFH model also prefers a lower mass-weighted stellar age of $t_{\star} = 5^{+61}_{-4}\, \rm Myr$, albeit with a substantial uncertainty. This SFH model returns a low best-fitting stellar mass of $\log(M_{\star}/\mathrm{M_{\odot}})=8.23^{+1.14}_{-0.21}$, but with an uncertainty encapsulating the potential for significant past episodes of star formation.  We adopt the outcome of SED modeling with the continuity bursty SFH model (see Table.~\ref{tab:bagpipes}), as it provides a realistic insight into the uncertainties when dealing with photometric and spectroscopic data covering just the rest-frame UV.

\begin{deluxetable*}{lcccccc}
\label{tab:bagpipes}
\caption{Properties derived from spectro-photometric modeling of PAN-z14-1.}
\tablehead{ID & $\log(M_{\star}/ \mathrm{M_{\odot}})$ & $t_{\star}$ & $A_{\rm V}$ & $\rm SFR_{10}$ & $\log(U)$ & $Z$ \\ 
 & & /Myr & /mag & /$\rm M_{\odot}\, yr^{-1}$ & & /$\rm Z_{\odot}$  }

\startdata
   PAN-z14-1 & $8.23^{+1.14}_{-0.21}$ & $5^{+61}_{-4}$ & $0.06^{+0.03}_{-0.03}$ & $4.8^{+13.6}_{-4.8}$ & $-1.4^{+0.9}_{-1.1}$ & $0.04^{+0.61}_{-0.03}$ \\
\enddata
\tablecomments{The modeling was performed with \textsc{bagpipes} assuming a non-parametric, continuity bursty star-formation history \citep{tacchella2022}. The table lists the best-fitting values of stellar mass, mass-weighted stellar age, $V$-band dust attenuation, star-formation rate over the last $10\, \rm Myr$, ionization parameter and metallicity.}
\end{deluxetable*}

\subsection{Emission lines}
Although the NIRSpec PRISM spectrum of PAN-z14-1 does not show any obvious emission lines, to quantify the significance and implications of any non detections we performed direct integration of the spectrum at the positions of common emission lines. We integrated the continuum-subtracted spectrum at the expected positions of the \niv, \civ, \heii $+$ \oiiiuv, \ciii, \mgii\, and \nev\, emission lines with a window of width $4\sigma_{\rm R}(\lambda)$, where the continuum is based on the best-fitting \textsc{bagpipes} model and $\sigma_{\rm R}(\lambda)$ is the expected standard deviation of a Gaussian emission line determined by the variable spectral resolution of the NIRSpec PRISM.
None of these lines are detected at $>2\sigma$ significance, and so we quote the resulting 
$2\sigma$ upper limits on line flux and rest-frame equivalent width in  Table~\ref{tab:line_measurements}.

\begin{deluxetable}{lcc}
\tablewidth{\columnwidth}
\label{tab:line_measurements}
\caption{Emission-line flux limits for PAN-z14-1.}
\tablehead{Emission line & Flux  & EW$_0$ \\
 & /$10^{-19}\, \rm erg \, s^{-1}\, cm^{-2}$ & $/\rm \AA$}

\startdata
  \niv & $<5.5$ & $<14$ \\
  \civ & $<5.2$ & $<14$  \\
  \heii $+$ \oiiiuv  & $<4.2$ & $<13$  \\
  \ciii & $<2.8$ & $<12$  \\
  \mgii & $<1.9$ & $<19$  \\
  \nev & $<2.6$ & $<39$  \\
\enddata
\tablecomments{The $2\sigma$ upper limits on the fluxes and equivalent widths were determined via direct integration of the spectrum.}
\end{deluxetable}

As a result of this analysis we find that the signal-to-noise ratio of the spectrum is sufficiently high to eliminate the possibility that PAN-z14-1 has rest-frame UV emission lines as strong as those observed in the $z=12.34$ galaxy GHZ2 \citep[e.g. $\mathrm{EW_{CIV}}\simeq46\, \rm \AA$,][]{castellano2024} or the $z=14.44$ galaxy MoM-z14 \citep[e.g. $\mathrm{EW_{CIV}}\simeq44\, \rm \AA$,][]{naidu2025}. Instead, PAN-z14-1 appears to exhibit the more ``normal'' emission-line strengths of luminous star-forming galaxies \citep[e.g.][]{carniani2024,robertsborsani2025}. In Section~\ref{sec:line_discussion} we discuss the implications of this for the galaxy population at $z>10$.

\subsection{Morphology}
\begin{figure}[ht!]
\centering
\includegraphics[width=\columnwidth]
{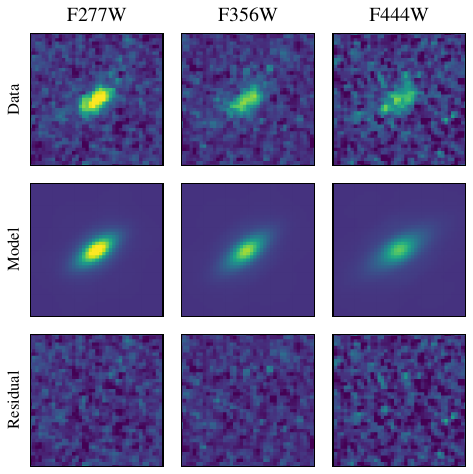}
\caption{The morphological fits to PAN-z14-1. The top row shows postage stamps of PAN-z14-1 ($1.2\times 1.2\, \rm arcsec$). The middle and bottom rows show the best-fitting \textsc{galfit} model and corresponding residuals, respectively.}
\label{fig:galfit}
\end{figure}

PAN-z14-1 is clearly extended across all the NIRCam imaging that it is detected in. In order to measure the size, we analyzed the F200W, F277W, F356W and F444W images using the 2D profile-fitting code \textsc{galfit} \citep{peng2002}. We show examples of the morphology fitting results in Fig. \ref{fig:galfit} for the F277W, F356W and F444W filters.

The fits in the F277W, F356W and F444W filters are all consistent with a disk-like morphology (i.e., S\`ersic index $n\simeq1$) and return broadly consistent effective radii (semi-major axis) in the range $450-550$ pc (adopting $z_{\rm spec}=13.53$). The fits to the F200W filter do not converge on a consistent value for the S\`ersic index, but return an effective radius of $348$ pc with a fixed value of $n=1$. Given the F277W displays the highest signal-to-noise, we adopt the measurement from this filter as our fiducial size measurement with an effective semi-major axis of $\mathrm{443\pm15\, pc}$.

Given the axial ratio of $\simeq4$, the corresponding circularized half-light radius ($r_{\rm c}$) is $233\pm10$ pc. The circularized UV half-light radius of PAN-z14-1 is therefore significantly larger than that of GHZ2 ($r_{\rm c}=105\pm9$; \citealt{castellano2024}) and MoM-z14 ($\mathrm{r_{c}=74^{+0.15}_{-0.12}\,pc}$; \citealt{naidu2025}), but similar to that of JADES-GS-z14-0 ($\mathrm{r_{c}=260\pm20\,pc}$; \citealt{carniani2024}). We discuss the potential link between galaxy size/morphology and UV emission-line strength in Section 4.3. The large size and disk-like morphology of PAN-z14-1 makes it a compelling candidate for NIRSpec IFU or MIRI/MRS follow-up in the future.

\section{Discussion}
\label{sec:discussion}
\label{sec:redshift_comparison}
\subsection{The photometric selection of galaxies at $z\geq14$}
As previously mentioned, the spectroscopic redshift of PAN-z14-1 is somewhat lower ($\Delta z\simeq1$) than determined by any of the studies that originally reported it. Given that this discrepancy is at the $\gtrsim3\sigma$ level c.f. McLeod et al. (in prep) and the $\sim5\sigma$ level c.f. \citet{Weibel2025}, it is important to investigate potential causes.

In general, careful selection of bright and robust $z>9$ candidate samples with \textit{JWST} has led to successful spectroscopic confirmations (see, e.g., \citealt{castellano2023,castellano2024,napolitano2025,kokorev2025,donnan2025b}). That said, there have been some notable instances of a systematic over-estimation of photometric redshifts \citep[e.g.][]{serjeant2023,arrabalharo2023b}, with one possible cause identified as the failure to account for the impact of the Lyman-$\alpha$ damping wing \citep{hainline2024}. 

\begin{figure}[ht!]
\centering
\includegraphics[width=\columnwidth]
{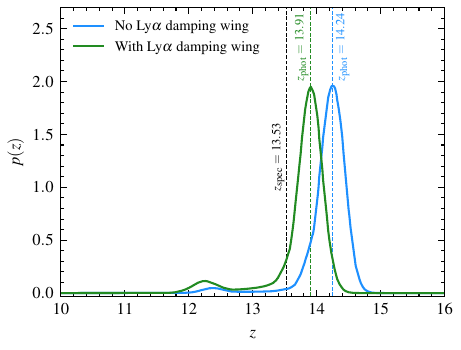}
\caption{The $p(z)$ distribution from SED fitting the NIRCam photometry of PAN-z14-1 using the code from \citet{mclure2011}, both with (green) and without (blue) accounting for the Lyman-$\alpha$ damping wing. The position of the spectroscopic redshift ($z_{\rm spec}=13.53$) is shown by the dashed black line. Once the Lyman-$\alpha$ damping wing is accounted for, the spectroscopic redshift lies within the $p(z)$ peak centered at $z_{\rm phot}=13.91^{+0.18}_{-0.19}$. For clarity, the plot is shown at $z>10$ and the integrated $p(z)$ at $z<10$ is negligible.}
\label{fig:p_z}
\end{figure}

To investigate this issue we have compared the best-fitting photometric redshift for PAN-z14-1 both with and without incorporating the Lyman-$\alpha$ damping wing, using an updated version of the code described in \citet{mclure2011}. 
Without the Lyman-$\alpha$ damping wing, we measure $z_{\rm phot}=14.24^{+0.20}_{-0.19}$, whereas with the damping wing included we estimate $z_{\rm phot}=13.91^{+0.18}_{-0.19}$ (see Fig.~\ref{fig:p_z}). In this case, the discrepancy between the photometric and spectroscopic redshift is reduced to the $2\sigma$ level, and the spectroscopic redshift of $z_{\rm spec}=13.53^{+0.05}_{-0.06}$ now resides within the broad $p(z)$ peak at $z\simeq14$.

Another key issue in determining an accurate photometric redshift for PAN-z14-1 is the position of the Lyman-$\alpha$ break which, at $z_{\rm spec}=13.53$, sits exactly on the short-wavelength edge of the F200W filter. As can be seen in Fig. 3, the measured F200W NIRCam photometry lies 10-15\% lower than both the NIRSpec spectrum and the best-fitting SED models. Tests reveal that this small flux offset alone is sufficient to move the best-fitting photometric redshift from $z_{\rm phot}\simeq 13.9$ to $z_{\rm phot}\simeq 13.5$. This suggests that the original $\Delta z\simeq 1$ over-estimate of the photometric redshift of PAN-z14-1 can be explained by a combination of a failure to account for the Lyman-damping wing and a random $\simeq 1\sigma$ downward scattering of the F200W flux. The example of PAN-z14-1 reinforces the lesson that the width of the NIRCam broad-band filters can make the photometric redshifts of high-redshift candidates uncertain at the $\Delta z\simeq 0.5$ level, and that this effect will be exacerbated if the Lyman-$\alpha$ break lies in the gaps between the broad-band filters.

One obvious method for addressing this issue is to improve the spectral resolution of the NIRCam data with the addition of medium-band imaging. Indeed, in the case of PAN-z14-1, the availability of deep F182M imaging would have been ideal. However, in reality, there are a couple of reasons to suggest that medium-band imaging is perhaps not a viable solution. The first problem is the depth required to make a significant impact. Ideally, one would like medium-band imaging of comparable depth to the broad-band imaging. Unfortunately, given the obvious differences in filter widths, this requires the investment of very large amounts of observing time. For example, current Cycle-4 medium-band programs in key extra-galactic fields (e.g., MINERVA, SPAM) fail to reach the required depths (e.g., the F210M imaging in MINERVA UDS is $\simeq 0.8$\,mag shallower than the F200W imaging). The second problem is the low number-density of $z\geq14$ galaxy candidates, which means that the medium-band imaging would not only have to be deep, but also cover large survey areas to be effective. Together, this suggests that robust pre-selection from NIRCam broad-band imaging, combined with efficient NIRSpec PRISM follow-up observations, remains the most efficient way to verify the redshifts of $z\geq 14$ galaxies.

There have also been a number of very luminous $z\geq13$ candidates that have multi-epoch imaging. For example, as noted by \citet{castellano2025,Weibel2025}, there is a $z\sim18$ candidate in the GLASS imaging of the Abell2744 cluster field that initially appears to be highly robust. However, this transient is absent in the GLASS epoch 1 imaging and then falls into a chip gap in the SW filters in epoch 2, creating the false appearance of a Lyman-$\alpha$ break at $z\sim18$. 
The recently reported, highly luminous ($M_{\rm UV}\simeq-21.2$), candidate at $z_{\rm phot}=13.7$ identified in the BEACON imaging \citep{Weibel2025,zhang2026} suffers from a similar issue. Although extremely luminous in the BEACON imaging, this source is entirely absent in earlier imaging obtained by the CEERS program. This suggests that this source is again likely to be a transient, rather than a galaxy at $z\simeq13.7$. 
In addition to contamination from low-$z$ interlopers, these two examples highlight additional sources of potential contamination when selecting of high-$z$ galaxies and demonstrate that careful consideration is required when combining imaging from multiple epochs.

Although we refrain from making definitive conclusions regarding the UV LF at $z\geq14$ from a single source, the now three spectroscopically confirmed galaxies at $z\geq13.5$ and $M_{\rm UV}<-20$, are consistent with measurements of the UV LF at $z\geq14$ from \citet{donnan2024, mcleod2024, Weibel2025} and D. J. McLeod et al. (in prep.). These measurements suggest a more rapid evolution in the UV LF at $z\geq14$ than is observed at $z=9-14$. Using the UV LF model from \citet{donnan2025} at $z=14$ (which matches these observational measurements), we predict $\simeq3$ galaxies at $z\geq13.5$ and $M_{\rm UV}<-20$ from the majority of the usable area of deep NIRCam imaging ($\simeq2300\, \rm sq.\, arcmin$) in remarkable agreement to the three which are spectroscopically confirmed. 

\subsection{Inferring the physical properties of galaxies at $z>10$ with SED modeling}
As noted in Section~\ref{sec:SED_fitting}, the inferred stellar mass of PAN-z14-1 is 
highly dependent on the assumed SFH, varying by $\simeq1.3\, \rm dex$. 
These differences arise due to the lack of constraints on the presence of a potential older stellar population, with the NIRSpec and NIRCam data probing only the rest-frame UV at $z>12$. 
Without longer-wavelength data, we cannot exclude the possibility of the higher stellar masses implied by the delayed-$\tau$ and DPL models. This effect is sometimes termed ``outshining'' and has been observed in galaxies at various redshifts \citep{papovich2001,conroy2010,Pforr2012}. This demonstrates the significant uncertainty in inferring stellar masses and star-formation histories at $z>10$ with NIRCam and NIRSpec data alone. 

It is worth noting that the implied stellar age returned by both the DPL and delayed-$\tau$ SFH models is $>100\, \rm Myr$. This implies that the galaxy formed at $z\sim30$, with a star-formation rate (and therefore UV magnitude) comparable to the observed $z_{\rm spec}=13.53$ value. This would imply the existence of highly luminous ($M_{\rm UV}\simeq-20.6$) galaxies up to $z\sim30$, which have not been observed in the major \textit{JWST} imaging programs (e.g. PRIMER, COSMOS-WEB, CEERS, JADES). Although seemingly unlikely, the low number density and small cosmological volume per unit redshift at $z\sim30$, currently makes this scenario hard to rule out with 100\% confidence.

Moreover, the recent detection of stellar continuum using $7.7\mu$m MIRI imaging of GS-z14-0 \citep{helton2025}, potentially suggest a significant stellar mass of $\log(M_{\star}/\mathrm{M_{\odot}})\simeq9.4$ in this galaxy at $z=14.18$. However, \citet{helton2025} disfavor this high-stellar mass given the detection of far-IR emission lines with ALMA \citep{carniani2025} and rest-frame optical lines with MIRI \citep{helton2025b}.

Alternatively, the very young mass-weighted stellar age of $t_{\star}=5\, \rm Myr$ implied by the non-parametric SFH implies a low mass-to-light ratio. This is consistent with what has been observed in other galaxies at $z\geq10$ \citep[e.g.,][]{curtislake2022,tang2025,donnan2025b,naidu2025}. This young stellar age is also consistent with the model predictions of \citet{donnan2025} which show that the slow evolution of the UV LF at $z=6-13$ can be explained with decreasing mass-to-light ratios driven by ever younger stellar ages towards higher redshifts. This model predicts that at $z>13$ the average stellar age is $t\simeq5\, \rm Myr$, consistent with the age implied by the non-parametric SFH. Overall,  however, the current data are unable to strongly constrain the SFH and stellar mass of PAN-z14-1.

\subsection{Are there two distinct galaxy populations at $z>10$?}
\label{sec:line_discussion}
The diversity in rest-frame UV emission-line strength displayed by $z\geq10$ galaxies has been noted in several previous studies. Some sources have been shown to exhibit weak/no carbon and nitrogen lines \citep[e.g.][]{curtislake2022,arrabalharo2023,carniani2024}, whereas others have yielded strong \ciii, \civ, and \niv\,emission lines \citep[e.g.][]{bunker2023b,castellano2024,naidu2025}. For example, the bright ($M_{\rm UV}=-20.5$) $z_{\rm spec}=12.34$ galaxy GHZ2 has strong \civ\, with a rest-frame equivalent width of $\simeq46\, \rm \AA$. These strong lines have been used to determine the carbon and nitrogen abundances relative to oxygen, finding elevated nitrogen abundances and sub-solar carbon abundance \citep[e.g.,][]{cameron2023}. 

\begin{figure}[ht!]
\centering
\includegraphics[width=\columnwidth]
{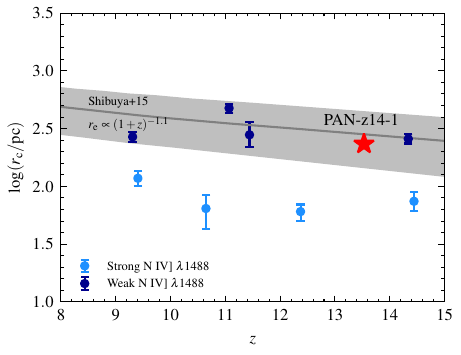}
\caption{A comparison of the size and redshift of PAN-z14-1 to other notable sources at $z>8$, split by the strength of their \niv\ emission.
In dark blue we show weak \niv\ emitters: Gz9p3 \citep{boyett2024}, CEERS2-588, Maisie's galaxy \citep{arrabalharo2023} and GS-z14-0 \citep{carniani2024}. In light blue we show strong \niv\ emitters: GNz9p4 \citep{schaerer2024}, GN-z11 \citep{bunker2023b}, GHZ2 \citep{castellano2024} and MoM-z14 \citep{naidu2025}. The gray region shows the size-redshift relation for `normal' star-forming galaxies from \citep{shibuya2015}. The relatively large size of PAN-z14-1 is consistent with its weak rest-frame UV emission lines.}
\label{fig:size-z}
\end{figure}

Recently, \citet{harikane2025} noted a possible connection between between galaxy morphology and rest-frame UV emission-line strength, with stronger lines found in more compact galaxies. Key examples of compact galaxies ($r_{\rm c}\lesssim100\, \rm pc$) with strong emission lines include GN-z11 \citep{bunker2023b}, GHZ2 \citep{castellano2024} and MoM-z14 \citep{naidu2025}. This relationship was also investigated by \citet{robertsborsani2025}, who found that while strong line emitters tend to be compact, weak line emitters span a wide range of physical sizes. To place PAN-z14-1 into this context, in Fig.~\ref{fig:size-z} we compare its size with several other well-studied sources at $z>8$.

The relatively large half-light radius of PAN-z14-1 is consistent with this correlation between physical size and the strength of rest-frame UV emission lines, as the upper limits on these lines rule out strong \niv, \civ\ and \ciii\ emission. The physical cause for this potential correlation remains unclear, although several suggestions have been advanced. 

One possibility is that there are two distinct populations of luminous galaxies at $z>10$, separated in terms of their star-formation surface density \citep{schaerer2024}. Using the SFR (averaged over $10\, \rm Myr$) noted in Table~\ref{tab:bagpipes}, PAN-z14-1 has a star-formation surface density of $\Sigma_{\rm SFR}=25^{+70}_{-24}\, \rm M_{\odot}\, yr^{-1}\, kpc^{-2}$. Therefore, although it exhibits a similar SFR to the significantly more compact luminous galaxies observed at $z>10$, due to its large physical size, PAN-z14-1 has a significantly smaller $\Sigma_{\rm SFR}$. For example, the compact and strong-emitter MoM-z14 has a similar SFR, but a significantly enhanced star-formation surface density $\Sigma_{\rm SFR}=233^{+107}_{-107}\, \rm M_{\odot}\, yr^{-1}\, kpc^{-2}$ \citep{naidu2025}. This could suggest that there are distinct populations of galaxies at $z>10$ where some undergo star-formation in very compact areas (e.g., GN-z11, GHZ2, MoM-z14), leading to elevated N/O, whereas others are forming stars over much larger areas (e.g., PAN-z14-1, GS-z14-0), lacking the compactness to generate sufficient ionizing radiation to produce strong UV emission lines. 

Alternatively, it has also been suggested that, rather than distinct evolutionary pathways, the strong and weak line emitters may 
simply represent different phases of the same evolutionary path \citep{robertsborsani2025}. Here, areas undergoing more intense episodes of star formation ``outshine'' the rest of the galaxy, leading to the apparently small physical sizes. It is also possible that given their compact sizes, these strong-line emitters may also host AGN. In particular this has been suggested for GN-z11 \citep{maiolino2024} although this is currently debated \citep{bunker2023b}.

\subsection{Similarity to GS-z14-0}
\label{sec:gsz14}
\begin{figure}[ht!]
\centering
\includegraphics[width=\columnwidth]
{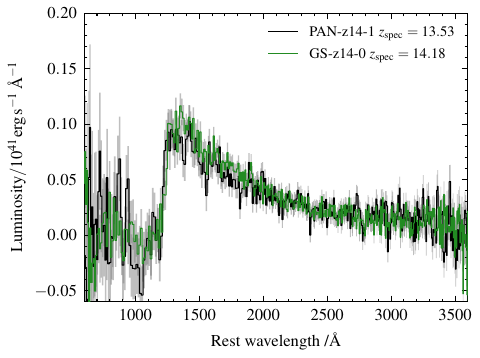}
\caption{Spectroscopic comparison of the rest-frame spectra of PAN-z14-1 and GS-z14-0 \citep{carniani2024}. These two galaxies are remarkably similar in terms of their UV luminosities, spectral slopes, physical sizes and lack of strong emission lines.}
\label{fig:spec_comparison}
\end{figure}

As discussed above, in terms of UV luminosity, spectral slope, emission-line strength and size, PAN-z14-1 is remarkably similar to the $z_{\rm spec}=14.18$ galaxy GS-z14-0 \citep{carniani2024}. Both galaxies exhibit similar UV luminosities ($M_{\rm UV}=-20.6\pm0.2$ for PAN-z14-1,  $M_{\rm UV}=-20.8\pm0.2$ for GS-z14-0), UV-continuum slopes ($\beta\simeq-2.3$), physical sizes ($r_{\rm c}\simeq250\, \rm pc$) and a lack of strong UV emission lines. The only UV emission line seen in GS-z14-0 is a tentative $3.6\sigma$ \ciii\ detection, with an equivalent width of $\simeq8\, \rm \AA$. We fail to achieve a formally significant detection of \ciii\ in PAN-z14-1, with a $2\sigma$ upper limit for the equivalent width of $12\, \rm \AA$. However, this is consistent with the line strength reported for GS-z14-0, the spectrum of which has a slightly higher signal-to-noise. In fact, our spectrum of PAN-z14-1 does arguably show a tentative feature at $\lambda\simeq2.75\, \mu$m, which could correspond to \ciii\ with an equivalent with of $\simeq6\, \rm \AA$ at a consistent redshift of $z\simeq13.45$, but we currently lack the signal-to-noise ratio required to confirm this detection. To further illustrate the similarity of these two galaxies, we directly compare their spectra in Fig.~\ref{fig:spec_comparison}. The similarity is obvious, as is the contrast to the compact, strong UV line emitter MoM-z14 at $z_{\rm spec}=14.44$ \citep{naidu2025}.

GS-z14-0 has also been observed with MIRI LRS \citep{helton2025b} revealing a $\sim4\sigma$ detection of H$\alpha$ with an exposure time of $\sim34\, \rm hours$. This confirms a star-formation rate of $\simeq10\, \rm M_{\odot}\, yr^{-1}$ in GS-z14-0 similar to what is expected in PAN-z14-1. If the similarity between GS-z14-0 and PAN-z14-1 continues to hold, this implies the significantly shorter planned MIRI LRS observations for PAN-z14-1 (GO 9425, PI: Schouws) will fail to detect H$\alpha$ with an exposure time of only $\sim11\, \rm hours$. However, the detection of \oiiiboth\, is more likely given that it is significantly brighter and detected with a greater signal-to-noise in GS-z14-0 than H$\alpha$ ($\simeq11\sigma$).

\section{Conclusion}
\label{sec:conclusion}
In this work we have presented \textit{JWST}/NIRSpec observations of a luminous and robustly-selected galaxy candidate at $z\geq14$. We have verified its redshift and discussed the physical properties of this galaxy. We summarize our conclusions below. 

\begin{itemize}
    \item We confirm the high-redshift nature of the robust $z\simeq14$ galaxy candidate PAN-z14-1 and measure a spectroscopic redshift of $z_{\rm spec}=13.53^{+0.05}_{-0.06}$ via modeling of the Lyman-$\alpha$ break. The moderate overestimation of our original photometric redshift for PAN-z14-1 ($z_{\rm phot}\simeq14.4$) is due to a combination of failing to account for the Lyman-$\alpha$ damping wing and a $\simeq 1\sigma$ downward scatter in the F200W flux. 

    \item PAN-z14-1 is one of the most UV luminous galaxies at $z>10$ discovered to date ($M_{\rm UV}=-20.6\pm0.2$), similar to GS-z14 and surpassed by only GN-z11. It exhibits a blue UV slope ($\beta=-2.26\pm0.08$). Through spectro-photometric modeling we derive a stellar mass of $\log(M_{\star}/\mathrm{M_{\odot}})=8.23^{+1.14}_{-0.21}$, with the large uncertainty driven by an inability to constrain previous star-formation episodes when fitting to 
    rest-frame UV data alone.

    \item The NIRSpec spectrum of PAN-z14-1 has sufficient signal-to-noise to rule-out the presence of strong UV emission lines, with $2\sigma$ upper limits of $\simeq14 \, \rm \AA$ on the equivalent widths of \niv, \civ, \heii\, and \ciii. This is consistent with recent suggestions that there may be two populations of luminous $z>10$ galaxies, with very different UV line strengths potentially explained by differences in star-formation surface density. In this scenario, the relatively large size of PAN-z14-1 ($r_{\rm c}\simeq233\ \rm pc$) makes it an excellent example of the population of more ``typical'' luminous star-forming galaxies, in contrast to a second recently-discovered population of extremely compact, strong UV line emitters.

\end{itemize}

This work reaffirms the ability of {\it JWST} NIRSpec PRISM data to efficiently confirm the redshifts and characterize the basic physical properties of $z>10$ galaxy candidates. In contrast to the population of extremely compact, strong UV line emitters at $z>10$, PAN-z14-1 joins a growing sample of luminous, spectroscopically confirmed, $z>10$ galaxies that exhibit more ``normal'' properties. The discovery of objects like PAN-z14-1 further demonstrates
the diversity of physical properties displayed by galaxies at the highest redshifts.

\begin{acknowledgments}
This work is based on observations made with the NASA/ESA/CSA \textit{James Webb Space Telescope}, obtained at the Space Telescope Science Institute, which is operated by the Association of Universities for Research in Astronomy, Incorporated. The data were obtained from the Mikulski Archive for Space Telescopes (MAST) at the Space Telescope Science Institute. 
These observations are associated with program \#6954. DJM, JSD and LB acknowledge the support of the Royal Society through the award of a Royal Society University Research Professorship to JSD. RJM acknowledges the support of the Science and Technology Facilities Council. ACC and HL acknowledge support from a UKRI Frontier Research Grantee Grant (PI Carnall; grant reference EP/Y037065/1). PAH acknowledges support from NASA under award 80GSFC24M0006.

\end{acknowledgments}




%
\facilities{JWST (NIRCam and NIRSpec)}

\software{astropy \citep{2013A&A...558A..33A,2018AJ....156..123A,2022ApJ...935..167A},  
          Cloudy \citep{2013RMxAA..49..137F}, 
          Source Extractor \citep{1996A&AS..117..393B}
          }


\appendix
\section{Slit-loss correction}
\label{sec:apdx_slit_loss}

\begin{figure}[ht!]
\centering
\includegraphics[width=0.5\columnwidth]
{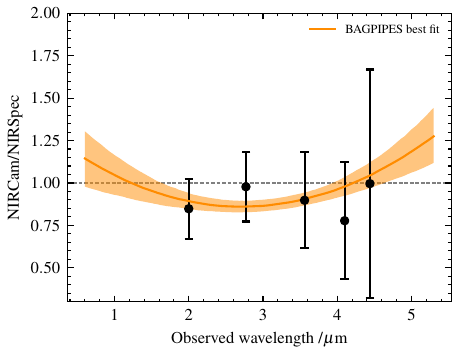}
\caption{The ratio of NIRCam flux to NIRSpec flux (integrated over the NIRCam filter band-passes) for all filters long-ward of the Lyman-$\alpha$ break (black data-points). The orange line shows the best-fitting second-order polynomial derived during SED fitting with \textsc{bagpipes} with the shaded region indicating the $1\sigma$ uncertainty.}
\label{fig:slit_loss_correction}
\end{figure}

As described in Section~\ref{sec:data_nirspec}, we correct the NIRSpec spectrum of PAN-z14-1 to match the measured NIRCam flux in each filter. This corrects for the loss of some flux that occurs when observing a source with NIRSpec as a result of instrumental effects as well as the small aperture size of the NIRSpec MSA slit. To correct for this, we fit a second-order polynomial to the ratio of NIRCam flux to the NIRSpec flux integrated over the NIRCam filter band-passes. These ratios are shown in Fig.~\ref{fig:slit_loss_correction} by the black data-points. The second-order polynomial is fitted as part of the SED fitting with \textsc{bagpipes}, therefore incorporating the associated uncertainty in this correction into the uncertainties of the derived physical properties of PAN-z14-1. The best fitting polynomial is shown by the orange line in Fig.~\ref{fig:slit_loss_correction}.

\section{Spectroscopic redshift measurement}
\label{sec:apdx_z_fit}
\begin{figure}[ht!]
\centering
\includegraphics[width=0.5\columnwidth]
{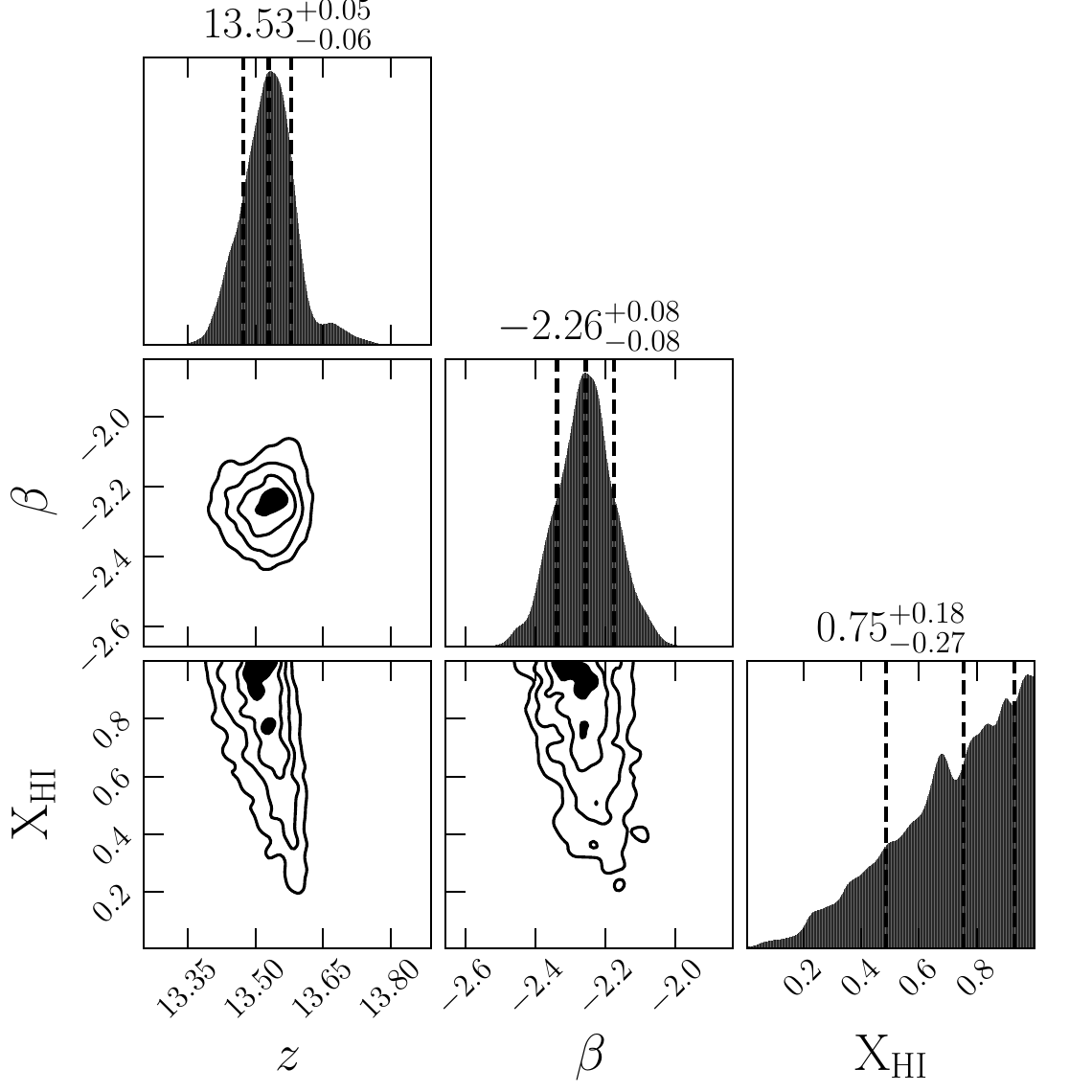}
\caption{2-D and 1-D posterior distributions for the three-parameter model to the Lyman-$\alpha$ break and UV-continuum following the method of \citet{cullen24}.}
\label{fig:z_fit_corner}
\end{figure}

In Section~\ref{sec:spec_z} we described the measurement of the spectroscopic redshift for PAN-z14-1 by fitting a model to the Lyman-$\alpha$ break as well as the UV-continuum following the method of 
\citet{cullen24}. This model contains three free parameters which are the redshift, $z$, UV-slope, $\beta$, and the fraction of neutral hydrogen gas in the IGM, $\rm X_{HI}$. The 2-D and 1-D posterior distributions for each parameter are shown in Fig.~\ref{fig:z_fit_corner}. 

\bibliography{z14}{}
\bibliographystyle{aasjournalv7}



\end{document}